\documentclass[aip,pof,reprint,amsmath,amssymb]{revtex4-1}

\usepackage{graphicx}
\begin{document}

\title{True, quasi and unstable Nambu-Goldstone modes of the two-dimensional Bose-Einstein condensed magnetoexcitons}

\author{S.A. Moskalenko,$^{1}$ M.A. Liberman,$^{2,3}$ D.W.Snoke,$^{4}$ E.V. Dumanov,$^{1}$
and S.S.Rusu,$^{1}$}

\email{michael.liberman@nordita.org}
\altaffiliation{}
\medskip
\affiliation{$^1$Institute of Applied Physics of the Academy of Sciences of Moldova, Academic Str. 5, Chisinau, MD2028, Republic of Moldova\\
$^2$Nordita, KTH Royal Institute of Technology and Stockholm University, Roslagstullsbacken 23, 10691 Stockholm, Sweden\\
$^3$Moscow Institute of Physics and Technology, Inststitutskii per. 9, Dolgoprudnyi, Moskovsk. obl.,141700, Russia\\$^4$Department of Physics and Astronomy, University of Pittsburgh,3941 O'Hara Street,
Pittsburgh, Pennsylvania 15260,USA} 

\date{\today}

\begin{abstract}
The collective elementary excitations of two-dimensional magnetoexcitons in
a Bose-Einstein condensate (BEC) with wave vector $\vec{k}=0$ were
investigated in the framework of the Bogoliubov theory of quasiaverages. The
Hamiltonian of the electrons and holes lying in the lowest Landau levels
(LLLs) contains supplementary interactions due to virtual quantum
transitions of the particles to the excited Landau levels (ELLs) and back.
As a result, the interaction between the magnetoexcitons with $\vec{k}=0$
does not vanish and their BEC becomes stable. The energy spectrum contains only one gapless, true Nambu-Goldstone(NG) mode of the second kind with dependence $\omega(k)\approx k^{2}$ at small values $k$ describing the optical-plasmon-type oscillations. There are two exciton-type branches corresponding to normal and abnormal Green's functions. Both modes are gapped with roton-type segments at intermediary values of the wave vectors and can be named as quasi-NG modes. The fourth branch is the acoustical plasmon-type mode with absolute instability in the region of small and intermediary values of the wave vectors. All branches have a saturation-type dependencies at great values of the wave vectors. The number and the kind of the true NG modes is in accordance with the number of the broken symmetry operators.
\end{abstract}


\maketitle

\section{Introduction}

A two-dimensional electron system in a strong perpendicular
magnetic field reveals fascinating phenomena such as the integer and
fractional quantum Hall effects. The discovery of the fractional
quantum Hall effect (FQHE) fundamentally changed the established concepts
about charged single-particle elementary excitations in solids [1, 2].

In this paper we study a coplanar electron-hole system with electrons
in a conduction band and holes in a valence band, both of which have Landau
levels in a strong perpendicular magnetic field. Earlier, this system was
studied in a series of papers [3-9] mostly dedicated to the theory of 2D
magnetoexcitons. This system bears some resemblance to the case of a bilayer electron system with half-filled lowest Landau levels in the conduction bands of each layer [10]. The coherent states of electrons in two layers happened to be equivalent to the BEC of the quantum Hall excitons [11] formed by electrons and holes in different layers. The system we are interested in has only one layer, with electrons
in conduction band and holes in the valence band of the same layer created
by optical excitation or by p-n doping injection (both of these methods can
be called \textquotedblleft pumping\textquotedblright). In the case of a single excited layer which we consider, the density
of excitons can be quite low, so that the electron Landau level and the
separate hole Landau level are each only slightly occupied, and Pauli
exclusion and phase space filling do not come in to play.
\section{Hamiltonian of the Bose-Einstein condensation of magnetoexcitons.}
\label{sec:2}
The effective Hamiltonian describing the interaction of electrons and holes lying
on the LLLs is
\begin{equation}
H=H_{Coul}+H_{Suppl}.
\end{equation}
Here $\hat{H}_{Coul}$ is the Hamiltonian of the Coulomb interaction of the electrons and holes lying on their LLLs:
\begin{equation}
\hat{H}_{Coul}=\dfrac{1}{2}\sum\limits_{\vec{Q}}W_{\vec{Q}}\left[ \hat{\rho}(\vec{Q})\hat{\rho}(-\vec{Q})-\hat{N}_{e}-\hat{N}_{h}\right] -\mu _{e}\hat{N}_{e}-\mu _{h}\hat{N}_{h},
\end{equation}
and $\hat{H}_{Suppl}$ is the supplementary indirect interactions between electrons and holes that appear
due to the simultaneous virtual quantum transitions of two particles from
the LLLs to excited Landau levels (ELLs) and their return back during the
Coulomb scattering processes. This interaction was deduced in Ref. [9]. It
has the form:
\begin{eqnarray}
H_{\text{suppl}} &=&\dfrac{1}{2}B_{i-i}\widehat{N}-\dfrac{1}{4N}%
\sum\limits_{Q}V(Q)\hat{\rho}(\vec{Q})\hat{\rho}(-\vec{Q})-  \notag \\
&&-\dfrac{1}{4N}\sum\limits_{Q}U(Q)\hat{D}(\vec{Q})\hat{D}(-\vec{Q}).
\end{eqnarray}
The coefficients $B_{i-i}$, $V(Q)$ and $U(Q)$ are proportional to the small parameter $r=I_{l}/(\hbar\omega_{c})<1$, where $I_{l}\cong\sqrt{B}$ is the ionization potential of the magnetoexciton, $\hbar\omega_{c}\sim B$ is the cyclotron energy, and $B$ is the magnetic field strength [9]. The degeneracy of the Landau level $N$ equals $S/(2\pi l^{2})$, where $l$ is the magnetic length $l^{2}=c\hbar/eB$ and $S$ is the surface layer area.
Here $\hat{\rho}(\vec{Q})$ are the density fluctuation operators expressed
through the electron $\hat{\rho}_{e}(\vec{Q})$ and hole $\hat{\rho}_{h}(\vec{%
Q})$ density operators as follows
\begin{gather}
\widehat{\rho }_{e}(\overrightarrow{Q})=\sum\limits_{t}e^{iQ_{y}tl^{2}}a_{t-%
\dfrac{Q_{x}}{2}}^{\dag }a_{t+\dfrac{Q_{x}}{2}};  \notag \\
\widehat{\rho }_{h}(\overrightarrow{Q})=\sum\limits_{t}e^{iQ_{y}tl^{2}}b_{t+%
\dfrac{Q_{x}}{2}}^{\dag }b_{t-\dfrac{Q_{x}}{2}},  \notag \\
\hat{\rho}(\vec{Q})=\hat{\rho}_{e}(\vec{Q})-\hat{\rho}_{h}(-\vec{Q}); \\
\hat{D}(\vec{Q})=\hat{\rho}_{e}(\vec{Q})+\hat{\rho}_{h}(-\vec{Q});  \notag \\
\hat{N}_{e}=\widehat{\rho }_{e}(0);\text{\ \ }\hat{N}_{h}=\widehat{\rho }%
_{h}(0);\text{\ \ }\hat{N}=\hat{N}_{e}+\hat{N}_{h};\text{\ \ }  \notag \\
W_{\vec{Q}}=\dfrac{2\pi e^{2}}{\varepsilon _{0}S\left\vert \vec{Q}%
\right\vert }e^{-Q^{2}l^{2}/2}.  \notag
\end{gather}
The density operators are integral two-particle operators. They are
expressed through the single-particle creation and annihilation operators $%
a_{p}^{\dag },a_{p}$ for electrons and $b_{p}^{\dag },b_{p}$ for holes. $%
\varepsilon _{0}$ is the dielectric constant of the background; $\mu _{e}$
and $\mu _{h}$ are chemical potentials for electrons and holes. Coefficients $V(Q)$, $U(Q)$ and $B_{i-i}$ were deduced in [9, 12].

As discussed in previous papers [5-9, 13], the breaking of the gauge
symmetry of the Hamiltonian (1) can be achieved using the
Keldysh-Kozlov-Kopaev [14] method with the unitary transformation
\begin{equation}
\hat{T}(\sqrt{N_{ex}})=\exp [\sqrt{N_{ex}}(d^{\dag }(\vec{k})-d(\vec{k}))],
\end{equation}
where $d^{\dag }(k)$ and $d(k)$ are the creation and annihilation operators
of the magnetoexcitons and $\vec{k}$ is the wave vector of the condensate. In the electron-hole representation they are [5-9]:
\begin{gather}
d^{\dag }(\vec{P})=\dfrac{1}{\sqrt{N}}\sum\limits_{t}e^{-iP_{y}tl^{2}}a_{t+%
\dfrac{P_{x}}{2}}^{\dag }b_{-t+\dfrac{P_{x}}{2}}^{\dag }; \\
d(\vec{P})=\dfrac{1}{\sqrt{N}}\sum\limits_{t}e^{iP_{y}tl^{2}}b_{-t+\dfrac{%
P_{x}}{2}}a_{t+\dfrac{P_{x}}{2}};  \notag
\end{gather}
BEC of magnetoexcitons leads to the formation of a coherent macroscopic
state as a ground state of the system with wave function
\begin{equation}
\left\vert \psi _{g}(\bar{k})\right\rangle =\hat{T}(\sqrt{N_{ex}})\left\vert
0\right\rangle ;\text{\ \ \ }a_{p}\left\vert 0\right\rangle =b_{p}\left\vert
0\right\rangle =0.
\end{equation}
Here $\left\vert 0\right\rangle $ is the vacuum state for electrons and
holes. Though we kept arbitrary value of $\vec{k}$,
nevertheless our main goal is the BEC with $\vec{k}=0$ and we will consider
the interval $0.5>kl\geq 0$. The function (7) will be used in the Section 3
to calculate the averages values of the type $\left\langle D(\overrightarrow{%
Q})D(-\overrightarrow{Q})\right\rangle $. The transformed Hamiltonian (1)
looks like
\begin{equation}
\hat{\mathcal{H}}=T\left( \sqrt{N_{ex}}\right) HT^{\dag }\left( \sqrt{N_{ex}}%
\right) ,
\end{equation}
and is succeeded, as usual, by the Bogoliubov u-v transformations of the
single-particle Fermi operators
\begin{gather}
\alpha _{p}=\hat{T}\left( \sqrt{N_{ex}}\right) a_{p}\hat{T}^{\dag }\left(
\sqrt{N_{ex}}\right) =ua_{p}-\text{v}(p-\dfrac{k_{x}}{2})b_{k_{x}-p}^{\dag };
\notag \\
\alpha _{p}\left\vert \psi _{g}(\bar{k})\right\rangle =0; \\
\beta _{p}=\hat{T}\left( \sqrt{N_{ex}}\right) b_{p}\hat{T}^{\dag }\left(
\sqrt{N_{ex}}\right) =ub_{p}\text{+v}(\dfrac{k_{x}}{2}-p)a_{k_{x}-p}^{\dag };
\notag \\
\beta _{p}\left\vert \psi _{g}(\bar{k})\right\rangle =0.  \notag
\end{gather}
Here $\text{v}(t)=\text{v}e^{-ik_{y}tl^{2}}$. The coefficients $u=\cos{(\sqrt{2\pi l^{2}n_{ex}})}$ and $\text{v}=\sin{(\sqrt{2\pi l^{2}n_{ex}})}$ were determined in Ref.[8]. The equality $\text{v}=\sin{(\text{v})}$ can be satisfied only at $\text{v}<1$.

Along with this traditional way of transforming the expressions of the
starting Hamiltonian (1) and of the integral two-particle operators (4) and
(6), we will use the method proposed by Bogoliubov in his theory of
quasiaverages [15], remaining in the framework of the original operators.
The new variant is completely equivalent to the previous one, and both of
them can be used in different stages of the calculations. For example, the average values of products of two integral two-particle operators can be calculated using the wave function (7) and u-v transformations (9), whereas the equations of motion for the integral two-particle operator can be simply written in the starting representation.

The Hamiltonian (1) with the broken gauge symmetry is written below in the lowest
approximation of the theory of quasiavereges [15]. Side by side with the last term in (10) there are another smaller terms breaking the gauge symmetry. They were neglected. This approach is named as quasiaverages theory approximation [11].
\begin{gather}
\hat{\mathcal{H}}=\dfrac{1}{2}\sum\limits_{\vec{Q}}W_{\vec{Q}}\left[ \rho (%
\vec{Q})\rho (-\vec{Q})-\hat{N}_{e}-\hat{N}_{h}\right] -\mu _{e}\hat{N}_{e}-
\notag \\
-\mu _{h}\hat{N}_{h}+\dfrac{1}{2}B_{i-i}\widehat{N}-\dfrac{1}{4N}%
\sum\limits_{Q}V(Q)\hat{\rho}(\vec{Q})\hat{\rho}(-\vec{Q})-  \notag \\
-\dfrac{1}{4N}\sum\limits_{Q}U(Q)\hat{D}(\vec{Q})\hat{D}(-\vec{Q})- \\
-\tilde{\eta}\sqrt{N}\left( d^{\dag }(k)+d(k)\right) .  \notag
\end{gather}
Here the parameter $\tilde{\eta}$, which determines the breaking of the gauge symmetry, depends on the chemical potential $\mu $ and on the square root of the density similar to the case of weakly non-ideal Bose-gas considered by Bogoliubov [15]. In our case the density is proportional to the
filling factor $\nu =\text{v}^{2}$. We have:
\begin{eqnarray}
\mu =\mu _{e}+\mu _{h};\text{ }\bar{\mu}=\mu +I_{l};N_{ex}=\text{v}^{2}N;
\notag \\
\tilde{\eta} =(\tilde{E}_{ex}(k)-\mu )\text{v}=(E(k)-\Delta (k)-\bar{\mu})%
\text{v};  \notag \\
\tilde{E}_{ex}(k) =-I_{l}-\Delta (k)+E(k);  \notag \\
E(k) =2\sum\limits_{Q}W_{Q}\sin ^{2}\left( \dfrac{[K\times Q]_{z}l^{2}}{2}%
\right) ;
\end{eqnarray}
$\Delta (k)\sim r$ determines the shift of the ionization potential $I_{l}$ [9].

\section{Energy spectrum of collective elementary excitations}

The equations of motion for the integral two-particle
operators with wave vectors $\vec{P}\neq 0$ in the special case of BEC of
magnetoexcitons with $\vec{k}=0$ are
\begin{gather}
i\hbar \dfrac{d}{dt}d(\vec{P})= \notag \\
=(-\bar{\mu}+E(\vec{P})-\Delta (\vec{P}))d(\vec{P})-  \notag \\
-2i\sum\limits_{\vec{Q}}\tilde{W}(\vec{Q})sin\left( \dfrac{[\vec{P}\times
\vec{Q}]_{z}l^{2}}{2}\right) \times   \notag \\
\times \hat{\rho}(\vec{Q})d(\vec{P}-\vec{Q})-\dfrac{1}{N}\times   \notag \\
\times \sum\limits_{\vec{Q}}U(\vec{Q})cos\left( \dfrac{[\vec{P}\times \vec{Q}%
]_{z}l^{2}}{2}\right) \times   \notag \\
\times D(\vec{Q})d(\vec{P}-\vec{Q})+\tilde{\eta}\dfrac{D(\vec{P})}{\sqrt{N}};
\end{gather}
\begin{gather*}
i\hbar \dfrac{d}{dt}\hat{\rho}(\vec{P})= \\
=-i\sum\limits_{\vec{Q}}\tilde{W}(\vec{Q})sin\left( \dfrac{[\vec{P}\times
\vec{Q}]_{z}l^{2}}{2}\right) \times  \\
\times \lbrack \hat{\rho}(\vec{P}-\vec{Q})\hat{\rho}(\vec{Q})+\hat{\rho}(%
\vec{Q})\hat{\rho}(\vec{P}-\vec{Q})]+ \\
+\dfrac{i}{2N}\sum\limits_{\vec{Q}}U(\vec{Q})sin\left( \dfrac{[\vec{P}\times
\vec{Q}]_{z}l^{2}}{2}\right) \times  \\
\times \left[ D(\vec{P}-\vec{Q})D(\vec{Q})+D(\vec{Q})D(\vec{P}-\vec{Q})%
\right] ; \notag
\end{gather*}
\begin{gather*}
i\hbar \dfrac{d}{dt}\hat{D}(\vec{P})= \\
=-i\sum\limits_{\vec{Q}}\tilde{W}(\vec{Q})sin\left( \dfrac{[\vec{P}\times
\vec{Q}]_{z}l^{2}}{2}\right) \times  \\
\times \lbrack \hat{\rho}(\vec{Q})\hat{D}(\vec{P}-\vec{Q})+\hat{D}(\vec{P}-%
\vec{Q})\hat{\rho}(\vec{Q})]+ \\
+\dfrac{i}{2N}\sum\limits_{\vec{Q}}U(\vec{Q})sin\left( \dfrac{[\vec{P}\times
\vec{Q}]_{z}l^{2}}{2}\right) \times  \\
\times \lbrack \hat{D}(\vec{Q})\hat{\rho}(\vec{P}-\vec{Q})+\hat{\rho}(\vec{P}%
-\vec{Q})\hat{D}(\vec{Q})] \\
+2\tilde{\eta}\sqrt{N}\left[ d(\vec{P})-d^{\dag }(-\vec{P})\right] ;
\end{gather*}
where $\tilde{W}(\vec{Q})=W(\vec{Q})-V(\vec{Q})/2N$.

Following the equations of motion (12) we have introduces four interconnected Green's functions $G_{1j}(P,t)$ as well as their Fourier transformations $G_{1j}(p,\omega)$ at $T=0$ [16, 17] of the type $G_{1j}(P,t)=\left\langle \left\langle A_{j}(P,t);B(P,0)\right\rangle \right\rangle$, where $A_{1}(P,t)=d(P,t), A_{2}(P,t)=d^{\dag }(-P,t), A_{3}(P,t)=\rho(P,t)/\sqrt{N}, A_{4}(P,t)=D(P,t)/\sqrt{N}$ and an arbitrary operator $B(P,0)$ because its choice does not influence on the energy spectrum of the system. These Green's function can be named as one-operator Green's functions. At the right hand sides of the corresponding equations of motion there is a second generation of the two-operator Green's functions. A second generation of equations of motion derived for them containing in their right hand sides the three-operator Green's functions. Following the procedure proposed by Zubarev [17] the truncation of the chains of equations of motion was made and the three-operator Green's functions were presented as a products of one-operator Green's functions $G_{i,j}(P,\omega)$ multiplied by the averages of the type $\left\langle D(P)D(-P)\right\rangle$. The average values were calculated using the ground state wave function (7) and the u-v transformations (9). For example, it was obtained
\begin{equation}
\left\langle D(P)D(-P)\right\rangle =4u^{2}\text{v}^{2}N; \left\langle \rho(P)\rho(-P)\right\rangle =0.
\end{equation}
The Zubarev procedure is equivalent to a perturbation theory with a small parameter of the type $\text{v}^{2}(1-\text{v}^{2})$. The closed system of Dyson equations has the form
\begin{equation}
\sum\limits_{j=1}^{4}G_{1i}(\vec{P},\omega )\Sigma _{ij}(\vec{P},\omega
)=C_{1j};\text{ }j=1,2,3,4.
\end{equation}
There are 16 different components of the self-energy parts $\Sigma _{jk}(%
\vec{P},\omega )$ forming a $4\times 4$ matrix. Four diagonal self-energy parts are:
\begin{eqnarray}
\Sigma _{11}(\overrightarrow{P},\omega ) =-\Sigma _{22}^{\ast }(-\overrightarrow{P},-\omega )= \nonumber \\
=\hbar \omega +i\delta +\overline{\mu }-E(\overrightarrow{P})+\Delta (\overrightarrow{P})-\frac{\left\langle D(\overrightarrow{P})D(-\overrightarrow{P}%
)\right\rangle }{N^{2}}\times \nonumber \\
\times \underset{\overrightarrow{Q}\neq \overrightarrow{P}}{%
\sum }\frac{U^{2}(\overrightarrow{Q})\cos ^{2}\left( \frac{[\overrightarrow{P%
}\times \overrightarrow{Q}]_{z}l^{2}}{2}\right) }{\hbar \omega +i\delta +%
\overline{\mu }-E(\overrightarrow{P}-\overrightarrow{Q})+\Delta (%
\overrightarrow{P}-\overrightarrow{Q})};  \nonumber \\
\Sigma _{33}(\overrightarrow{P},\omega ) =\hbar \omega +i\delta -\frac{%
\left\langle D(\overrightarrow{P})D(-\overrightarrow{P})\right\rangle }{%
N^{2}(\hbar \omega +i\delta )}\times   \nonumber \\
\times \underset{\overrightarrow{Q}\neq \overrightarrow{P}}{\sum }U(%
\overrightarrow{Q})\left( U(-\overrightarrow{Q})-U(\overrightarrow{Q}-%
\overrightarrow{P})\right) \sin ^{2}\left( \frac{[\overrightarrow{P}\times
\overrightarrow{Q}]_{z}l^{2}}{2}\right) ; \\
\Sigma _{44}(\overrightarrow{P},\omega ) =\hbar \omega +i\delta -\frac{%
2\left\langle D(\overrightarrow{P})D(-\overrightarrow{P})\right\rangle }{%
N(\hbar \omega +i\delta )}\times   \nonumber \\
\times \underset{\overrightarrow{Q}\neq \overrightarrow{P}}{\sum }%
\widetilde{W}(\overrightarrow{Q})\left( U(\overrightarrow{Q}-\overrightarrow{%
P})-U(\overrightarrow{P})\right) \sin ^{2}\left( \frac{[\overrightarrow{P}%
\times \overrightarrow{Q}]_{z}l^{2}}{2}\right) +  \nonumber \\
+\frac{\left\langle D(\overrightarrow{P})D(-\overrightarrow{P}%
)\right\rangle }{N^{2}(\hbar \omega +i\delta )}\underset{\overrightarrow{Q}%
\neq \overrightarrow{P}}{\sum }U(\overrightarrow{Q})\left( U(\overrightarrow{%
P})-U(-\overrightarrow{Q})\right) \times   \nonumber \\
\times \sin ^{2}\left( \frac{[\overrightarrow{P}\times \overrightarrow{Q}%
]_{z}l^{2}}{2}\right) .  \nonumber
\end{eqnarray}
They contain real and imaginary parts as follows $\Sigma_{ij}(\vec{P},\omega)=\sigma_{ij}(\vec{P},\omega)+i\Gamma(\vec{P},\omega).$
The cumbersome dispersion equation for electron-hole collective excitations is expressed in general form by the
determinant equation
\begin{equation}
\det \left\vert \Sigma _{ij}(\vec{P},\omega )\right\vert =0.
\end{equation}
Due to the structure of the self-energy parts, it separates into two
independent equations. One of them concerns only the optical plasmon branch, corresponding to oscillations of the difference of electron and hole densities and has a simple form:
\begin{equation}
\Sigma _{33}(\vec{P},\omega )=0.
\end{equation}
It does not include at all the chemical potential $\bar{\mu}$ and the
quasiaverage constant $\tilde{\eta}$. But the average values depends also on the condensate wave function (7).

The solution of the equation (17) is
\begin{eqnarray}
& \left( \hbar \omega (\vec{P})\right) ^{2}=\dfrac{\left\langle \notag
D(\vec{P})D(-\vec{P})\right\rangle }{N^{2}}\sum\limits_{\vec{Q}}sin^{2}\left( \dfrac{[\vec{P}\times \vec{Q}]_{z}l^{2}}{2}\right)\times &  \\
& \times U(\vec{Q})\left( U(-\vec{Q})-U(\vec{Q}-\vec{P})\right)&
\end{eqnarray}
The right hand side of this expression at small values of $P$ has a
dependence $\left\vert P\right\vert ^{4}$ and tends to saturate at large
values of $P$. The optical plasmon branch $\hbar \omega _{OP}(P)$ has a
quadratic dispersion law in the long wavelength limit and saturation
dependence in the range of short wavelengths. Its concentration dependence
is of the type $\sqrt{\text{v}^{2}(1-\text{v}^{2})}$ what coincides with the
concentration dependencies for 3D and 2D plasmons. The obtained
dispersion law is represented in the Fig.1.

The second equation contains the self-energy parts $\Sigma _{11}$, $\Sigma _{22}$, $\Sigma _{44}$, $\Sigma
_{14}$, $\Sigma _{41}$, $\Sigma _{24}$ and $\Sigma _{42}$, which include the
both parameters $\bar{\mu}$ and $\tilde{\eta}$ and has the form
\begin{eqnarray}
&&\Sigma _{11}(\vec{P};\omega )\Sigma _{22}(\vec{P};\omega )\Sigma _{44}(%
\vec{P};\omega )-  \notag \\
&&-\Sigma _{41}(\vec{P};\omega )\Sigma _{22}(\vec{P};\omega )\Sigma _{14}(%
\vec{P};\omega )- \\
&&-\Sigma _{42}(\vec{P};\omega )\Sigma _{11}(\vec{P};\omega )\Sigma _{24}(%
\vec{P};\omega )=0.  \notag
\end{eqnarray}
The solutions of the equation (19) describe the exciton energy
and quasienergy branches arising due to the normal and abnormal Green's functions as well as the acoustical plasmon branch.
The ideal magnetoexciton gas can exist only in the case $\text{v}^{\text{2}%
}=0$, with an infinitesimal number of excitons, but without plasma at all. In this case the plasmon frequency is zero whereas the exciton of one exciton is possibly.

In the zero order approximation on the concentration, when $\nu=\text{v}^{2}=0$ the real parts $\sigma_{ii}(\vec{P},\omega)$ look as:
\begin{eqnarray}
\sigma _{11}(\overrightarrow{P},\omega )=-\sigma _{22}(-\overrightarrow{P}%
,-\omega )=  \notag \\
=\hbar \omega +\overline{\mu }-E(\overrightarrow{P})+\Delta (%
\overrightarrow{P});  \notag \\
\sigma _{33}(\overrightarrow{P},\omega )=\sigma _{44}(\overrightarrow{P}%
,\omega )=0; \\
\widetilde{\eta } =0; \mu +\Delta (0)=0  \notag
\end{eqnarray}
In this limit the energy needed to excite the magnetoexciton with $\vec{k}=0$ in the state with $\vec{P}\neq 0$ equals to $E(\vec{P})$, whereas the plasma oscillations do not exist at all. In this approximation the exciton branches of the spectrum are gapless and true NG modes. In the first order approximation on the small parameter $\text{v}^{2}(1-\text{v}^{2})$ we have a non-ideal Bose gas and the self-energy parts (15) contain terms linear and quadratic in the interaction constant $U(\vec{P})$. The last terms have the denominators with unknown frequency what increase the number of the solutions. They also can be taken into account also by the iteration method. The first step in this direction gives the following parts $\sigma_{ij}(\vec{P}, \omega)$
\begin{gather}
\sigma _{11}(\vec{P},\omega )=-\sigma _{22}(-\vec{P},-\omega )=\hbar \omega +\bar{\mu}-E(P)+\Delta (P);
\notag \\
\sigma _{41}(\vec{P},\omega )=-\tilde{\eta}+U(P)\dfrac{\left\langle
d(0)\right\rangle }{\sqrt{N}}; \\
\sigma _{42}(\vec{P},\omega )=\tilde{\eta}-U(-P)\dfrac{\left\langle
d(0)\right\rangle }{\sqrt{N}};  \notag \\
\sigma _{14}(\vec{P},\omega )=-2\tilde{\eta};\text{\ }\sigma _{24}(\vec{P%
},\omega )=2\tilde{\eta};
\sigma _{44}(\vec{P},\omega )=\hbar \omega ;  \notag
\end{gather}
Now the exciton branches of the spectrum transform from the true into the quasi-NG modes with the gaps in the point $\vec{P}=0$ as follows
\begin{gather}
\hbar \omega =\pm \sqrt{\left( \bar{\mu}-E(\vec{P})+\Delta (0)\right) ^{2}+4%
\tilde{\eta}\left( \tilde{\eta}-\dfrac{U(\vec{P})\left\langle
d(0)\right\rangle }{\sqrt{N}}\right) }
\end{gather}
In this approximation the acoustical plasmon branch continues to vanish. Taking into account the quadratic terms on the interaction constant $U(\vec{P})$ we have obtained the corrections to the exciton branches of the spectrum and the dispersion law for the acoustical plasmon mode.

The acoustical plasmon branch corresponding to oscillations of the total particle density has a dispersion law completely different from
the optical plasmon oscillations. It has an absolute instability beginning
with small values of wave vector going on up to the considerable value $%
Pl\approx 2$. In this range of wave vectors, the optical plasmons have
energies which do not exceed the activation energy $U(0)$. The supplementary Hamiltonian gives rise to general attraction in Hartree approximation, characterized by the coefficients $U(0)$. It plays the role of activation energy if one should like to overpass this attraction. The ground state of the system is unstable as regards the generation of the acoustical plasmons. It means that the system becomes a localized generator of the growing acoustical waves without their traveling through the system as in the case of convective instability.

In the case of 2D magnetoexcitons in the BEC state with small wave vector $kl<0.5$ described by the Hamiltonian (10), the both initial continuous symmetries are lost. It happened due to the presence of the term $\widetilde{\eta }(d^{\dag }(\overrightarrow{k})+d(\overrightarrow{k}))$ in the frame of the Bogoliubov theory of the quasiaverages. Nevertheless the energy of the ground state as well as the self-energy parts $\Sigma _{ij}(P,\omega )$ were calculated only in the simplest case of the condensate wave vector $\overrightarrow{k}=0$. These expressions can be relevant also for infinitesimal values of the modulus $|\textbf{k}|$ but with a well defined direction. In this case the symmetry of the ground state will be higher than that of the Hamiltonian (10), what coincides with the situation described by Georgi and Pais [18]. It is one possible explanation of the quasi-NG modes appearance in the case of exciton branches of the spectrum.
Another possible mechanism of the gapped modes appearance is the existence of the local gauge symmetry, the breaking of which leads to the Higgs effect [19]. The interaction of the electrons with the attached vortices gives rise to a gapped energy spectrum of the collective elementary excitations as was established in Ref.[20, 21].
The number of the NG modes in the system with many broken
continuous symmetries was determined by the Nielsen and Chadha [22] theorem.
It states that the number of the first-type NG modes $N_{I}$ being accounted
once and the number of the second type NG modes $N_{II}$ being accounted
twice equals or prevails the number of broken generators $N_{BG}$. It looks
as follows $N_{I}+2N_{II}\succeq N_{BG}$. In our case the optical plasmon
branch has the properties of the second-type NG modes. We have $N_{I}=0;$ $%
N_{II}=1$ and $N_{BG}=2$. It leads to the equality $2N_{II}=N_{BG}$.
The three branches of the energy spectrum are represented together in the Fig.1. One of them is a second-type Nambu-Goldstone(NG) mode describing the
optical plasmon-type excitations, the second branch is the first-type NG
mode with absolute instability describing the acoustical-type excitations
and the third branch is the quasi-NG mode describing the exciton-type
collective elementary excitations of the system. It is a gapless true NG mode in zero order approximation on the small parameter $\text{v}^2(1-\text{v}^2)$ and become gapped in the first order approximation due to the fact that the symmetry of the ground state with small condensate wave vector $\vec{k}$ is greater than the symmetry of the Hamiltonian.
\begin{figure}
\resizebox{0.48\textwidth}{!}{%
  \includegraphics{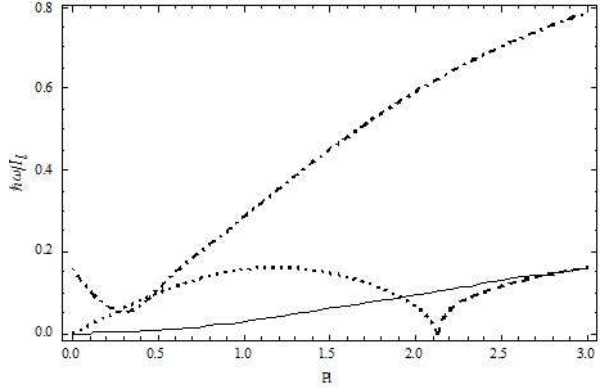}
}
\caption{Three branches of the collective elementary excitations: the
exciton-type quasi-NG mode with a gap in the point $pl=0$ (dash-dotted line); the second-type
NG mode describing the optical plasmons(solid line) and the first-type NG mode with
absolute instability (dotted line) describing the acoustical plasmons(dashed line).}
\label{fig:1}       
\end{figure}

\section{Conclusions}

The energy spectrum of the collective elementary excitations of a 2D
electron-hole system in a strong perpendicular magnetic field in the
state of Bose-Einstein condensation with wave vector $\vec{k}=0$ was
investigated in the framework of the Bogoliubov theory of quasiaverages.
The starting Hamiltonian describing the e-h system contains not only the
Coulomb interaction between the particles lying on the lowest Landau levels,
but also a supplementary interaction due to their virtual quantum
transitions from the LLLs to the excited Landau levels and back. This
supplementary interaction generates, after the averaging on the ground state
BCS-type wave function, direct Hartree-type terms with an attractive
character, exchange Fock-type terms giving rise to repulsion, and similar
terms arising after the Bogoliubov $u-v$ transformation. The interplay of
these three parameters gives rise to the resulting nonzero interaction
between the magnetoexcitons with wave vector $\vec{k}=0$ and to stability of
their BEC as regards the collops.

The separated electrons and holes remaining on their Landau orbits can take
part in the formation of magnetoexcitons as well as in collective plasma
oscillations. Such possibilities were not taken into consideration in the
theory of structureless bosons or in the case of Wannier-Mott excitons with
a rigid relative electron-hole motion structure without the possibility of
the intra-series excitations.

The energy spectrum of the collective elementary excitations consists of four branches. Two of them are excitonic-type branches(energy and quasienergy branches). The other two branches are the optical and acoustical plasmon branches.
We can say that results obtained in our system are similar to what was obtained in system of BEC of the quantum Hall excitons[11].
In these both models there is only one gapless Nambu-Goldstone mode between four branches of the energy spectrum. In both models the exciton branches of the spectrum are not gapless and differ from the NG modes. In our case the exciton energy and quasienergy branches corresponding to normal and abnormal Green's functions have a gaps in the point $P=0$, a roton-type segments in the region of intermediary wave vectors $Pl\sim1$ and saturation-type behaviors at great values of $Pl$.

We should like to thank the referee, whose remarks stimulated to essentially improve our paper.







\end{document}